\documentclass[dvips]{article}

\usepackage{icrc2011}

\title{Energy and $X_{max}$ reconstruction of hadron-initiated showers in surface arrays}

\newcommand{\etal}{\MakeLowercase{\textit{et al. }}} 
\shorttitle{G. Ros \etal Energy and $X_{max}$ of UHECRs from surface detectors}

\authors{G. Ros$^{1}$, G. A. Medina-Tanco$^{2}$, A. D. Supanitsky$^{3}$,  L. del Peral$^{1}$,  M. D. Rodr\'iguez Fr\'ias$^{1}$.}
\afiliations{$^1$Space and Astroparticle Group, Dpto. F\'isica, Universidad de Alcal\'a, (Spain).
\\ $^2$Instituto de Ciencias Nucleares, UNAM, (Mexico).
\\$^3$Instituto de Astronom\'ia y F\'isica del Espacio, CONICET-UBA, (Argentina).
}
\email{german.ros@uah.es}

\abstract{The current methods to determine the primary energy in surface arrays are different when dealing with hadron or photon initiated showers.
In this work, we adapt a method previously developed for photon-initiated showers to hadron primaries. We determine the Monte Carlo 
parametrizations that relate the surface energy estimator and the maximum of shower development for both, proton and Iron primaries. Using for each 
primary their own set of calibration curves, which is of course impossible in practice, we show that the energy could be inferred with a negligible 
bias and $12\%$ resolution. However, we show that a \textit{mixed} calibration could also be performed, including both type of primaries, 
such that the bias still remains low and the achieved resolution is around $15\%$. In addition, the method allows the simultaneous determination 
of $X_{max}$ in pure surface arrays with resolution better than $7\%$.}


\begin{document}
\maketitle

\section{Introduction}

In large ultra-high energy cosmic ray surface arrays the technique traditionally used to determine the primary energy
consists on the inference of the lateral distribution of particles of the extensive air shower (EAS) it produces into the atmosphere.
Scintillators (e.g., Volcano Ranch, AGASA, KASCADE) and water Cherenkov detectors (e.g., Haverah Park) have been mainly used 
for this purpose. The surface array allows the discrete sampling of the shower front at ground level and then, the lateral distribution 
is fitted assuming a certain functional form (LDF, lateral distribution function). Later, the signal inferred at an optimum distance 
is used as the energy estimator, which is related to the primary energy using Monte Carlo (MC) parametrizations. This optimum distance 
is traditionally fix for each experiment since it is assumed to be only dependent on the array spacing. However, recent studies 
suggest the convenience of calculating the optimum distance for each individual shower taking into account primary energy 
and direction \cite{GRos_ropt}. A special case from the experimental point of view is the Pierre Auger Observatory 
which pioneers the simultaneous use of water Cherenkov detectors and fluorescence telescopes. For these {\it hybrid} events, systematic 
errors in their energy estimate are greatly reduced \cite{AugerHybridSpectrumICRC09}.

These methods assume that primary is a proton and it is considered adequate for heavier primaries since 
the estimated energy for nuclei depends weakly on their mass number. On the other hand, the difference is significant when dealing with 
photon-initiated showers since the muonic component is much lower, shower development is affected by the geomagnetic field and the LPM effect
delays in average the first interaction. Thus, each experiment has followed a different method when searching for photon primaries.
Haverah Park and Auger use the muon density at ground \cite{HP_Photon} and MC parametrizations \cite{Auger_Photon} to infer 
the primary energy respectively, while in \cite{AGASA_Photon_1,AGASA_Photon_2}, AGASA data is directly compared to photon MC simulations. 

The method used by Auger \cite{Auger_Photon} was first proposed in \cite{Billoir}. The original idea was to adapt the energy reconstruction method
to the late development of photon showers and the key point is to rely explicitly on the development stage of the shower. The method was
originally applied to photon showers in an Auger-like array. An empirical parametrization between $S(1000)/E$ and $\Delta X = X_{ground}/cos \,\theta - X_{max}$
was found, where $S(1000)$ is the inferred signal at $1000$ meters from the shower axis, $E$ is the primary energy, $X_{max}$ is the maximum of 
shower development and $\theta$ is the zenith angle of the shower. The parametrization is almost independent of the primary energy due to the well-known 
universality of the electromagnetic component of EAS \cite{Lafebre,Kascade,Schmidt} and the small muon fraction in the photon-initiated showers. 

In this work, we show how to modify this method to be applicable to hadron-initiated showers where the muon component is significant, specially, in case
of water Cherenkov arrays which enhanced their contribution to the total measured signal. In this way it is possible to determine simultaneously the energy and 
$X_{max}$ of the shower from pure surface data. Alternatively, several surface parameters have been used to infer indirectly $X_{max}$, such as the rise 
time of the signals in the detectors and the azimuthal features of the time distributions \cite{AugerCompositionSDICRC09}.

\section{Simulations} \label{Sec:Simulations}

The simulation of the atmospheric showers is performed with the AIRES Monte Carlo program (version 2.8.4a) \cite{Aires} using QGSJET-II-03 
as the hadronic interaction model (HIM). The input primary energy goes from $log(E/eV)$ $=$ $19.0$ to $19.6$ in $0.1$ steps. Around $300$ events 
have been simulated per each energy and for both, proton and iron primaries. The zenith angle has been selected following a sine-cosine 
distribution from $30$ to $60$ degrees, while the azimuth angle is randomly distributed.

As it will be explained in Section \ref{Sec:Method}, only the reconstructed $S(1000)$ and zenith angle of the shower will be needed in the method 
proposed here. We simulate the reconstructed zenith angle by fluctuating the real one with a Gaussian whose standard deviation is $1^o$, 
a typical value of its uncertainty in surface arrays \cite{AugerAngularError, AGASAAngular, HiResAngular, HPAngular}. 

A realistic $S(1000)$ of the event could be obtained from
\begin{eqnarray}\label{CICHybrid}
E &=& A(S_{38})^B  \nonumber \\
S(1000)(\theta)&=&S_{38}\times\left[1+ Cx - Dx^2 \right]  
\end{eqnarray}
where $x = cos^2(\theta)-cos^2(38^o)$. $A$, $B$, $C$ and $D$ are constants obtained in \cite{Maris_Thesis} for QGSJet-II-03 and for iron and proton primaries. 
However, event by event fluctuations and reconstruction uncertainties are not taken into account if $S(1000)$ is directly estimated from  Eq. (\ref{CICHybrid}). 
Thus, we calculate the reconstructed $S(1000)$ from our own simulation of the detector response and reconstruction process, as it is explained next.

The simulation of the tank response and the fit of the LDF are performed using our own simulation program previously tested in 
\cite{GRos_ropt}. An infinite array whose unitary cell consists on a triangle with detectors separated 1.5 km is considered. The real core
is randomly located inside an elementary cell and the reconstructed one is determined fluctuating the real one with a Gaussian function whose
standard deviation depends on the primary energy and composition (more details en \cite{GRos_ropt}). 

The signal collected at each station for a given shower with a certain energy and zenith angle is set assuming a \textit{true} lateral 
distribution function like \cite{Maris_Thesis}
\begin{equation}\label{AugerLDF}
S(r)=S(1000)\times\left(\frac{r}{1000}\right) ^{-\beta}\times\left( \frac{r+700}{1000+700}\right)^{-\beta},
\end{equation}
where the distance to shower axis $r$ is in meters and $\beta(\theta,S(1000))$ is given by
\begin{equation}\label{BetaAugerLDF}
\beta(\theta,S(1000)) = \left\lbrace
  \begin{array}{l}
   a + b (sec\theta - 1), \hspace{0.3cm} if \hspace{0.2cm} sec\theta < 1.55\\
   a + b (sec\theta - 1) + f (sec\theta -1)^2,  \\ \hspace{2.5cm} if \hspace{0.2cm} sec\theta > 1.55\\   
  \end{array}
  \right\rbrace
\end{equation}
where $a=2.26 + 0.195 log(e)$, $b = -0.98$, $c = 0.37-0.51sec\theta +0.30sec^2\theta$, $d = 1.27-0.27sec\theta +0.08sec^2\theta$, $e=c(S(1000))^d$
and $f = -0.9$. $S(1000)$ in Eqs. (\ref{AugerLDF}) and (\ref{BetaAugerLDF}) is obtained from Eq. (\ref{CICHybrid}). Later, the signal assigned to each 
station is fluctuated using a Poissonian distribution whose mean is given by the \textit{true} LDF. We impose typical values for trigger condition and 
saturation. Saturated detectors are excluded from the LDF fit.

Next, the lateral distribution of particles is fitted using a functional form given by
\begin{equation}\label{AugerLDFFit}
\log S(r)=a_1+a_2\left[\log\left(\frac{r}{1000}\right)+\log\left(\frac{r+700}{1000+700}\right)\right],
\end{equation}
where the slope of the LDF and the normalization constant are determined in each fit considering the core position as fixed in the reconstructed
one. Finally, the reconstructed $S(1000)$ is determined as the interpolated value from the fit at $1000$ meters from the shower axis.

\section{Energy and $X_{max}$ reconstruction} \label{Sec:Method}

The method is based on MC parametrizations and, essentially, the energy of the primary and $X_{max}$ are iteratively determined. 
As mentioned, the only ingredients needed are the reconstructed zenith angle of the incoming shower, $\theta$, and the interpolated 
$S(1000)$ value from the LDF fit.

The iterative process should be started with an initial rough estimation of the energy. We arbitrarily use $10$, $20$ and $50$ EeV. 
The results do not depend on this choice. Next, $X_{max}$ is estimated using its average dependence on energy,
\begin{equation}\label{Xmax_vs_Energy}
X_{max} = q_0 + q_1 \times log_{10}(E_{prim})  \;\;\;\;\ g/cm^2,
\end{equation}
where $E_{prim}$ is in EeV. Next, using $S(1000)$ and $\theta$, primary energy could be obtained from
\begin{equation}\label{S1000_over_E}
S(1000)/E_{prim} = p_0 \times \frac {1 + \frac{\Delta X - 100}{p_1}} {1 + (\frac{\Delta X - 100 }{p_2})^2} \;\;\;\;\; VEM/EeV.
\end{equation}
Then, $E_{prim}$ obtained from Eq. (\ref{S1000_over_E}), is used in Eq. (\ref{Xmax_vs_Energy}) to get a new $X_{max}$ following an iterative 
process until convergence is achieved (we set the convergence in energy at $10^{-5}$). The convergence is always fast, only 3-6 steps 
are needed.


This method works properly in case of photon-initiated showers since $S(1000)/E_{prim}$ vs. $\Delta X$ shows an universal profile independently
of primary energy due to the low fraction of muons in photon showers as explained before. However, hadron primaries produce EAS with a significant 
muon component and, in addition,  the muon fraction depends on the primary energy. As consequence, if a global calibration curve is used, the 
inferred energy will show an energy dependent bias as shown in Fig. 1 
(black points).

\begin{figure}[!bt] \label{fig:EnergyBias_vs_logEreal_Proton}
\begin{center}
\includegraphics[width=8.5cm]{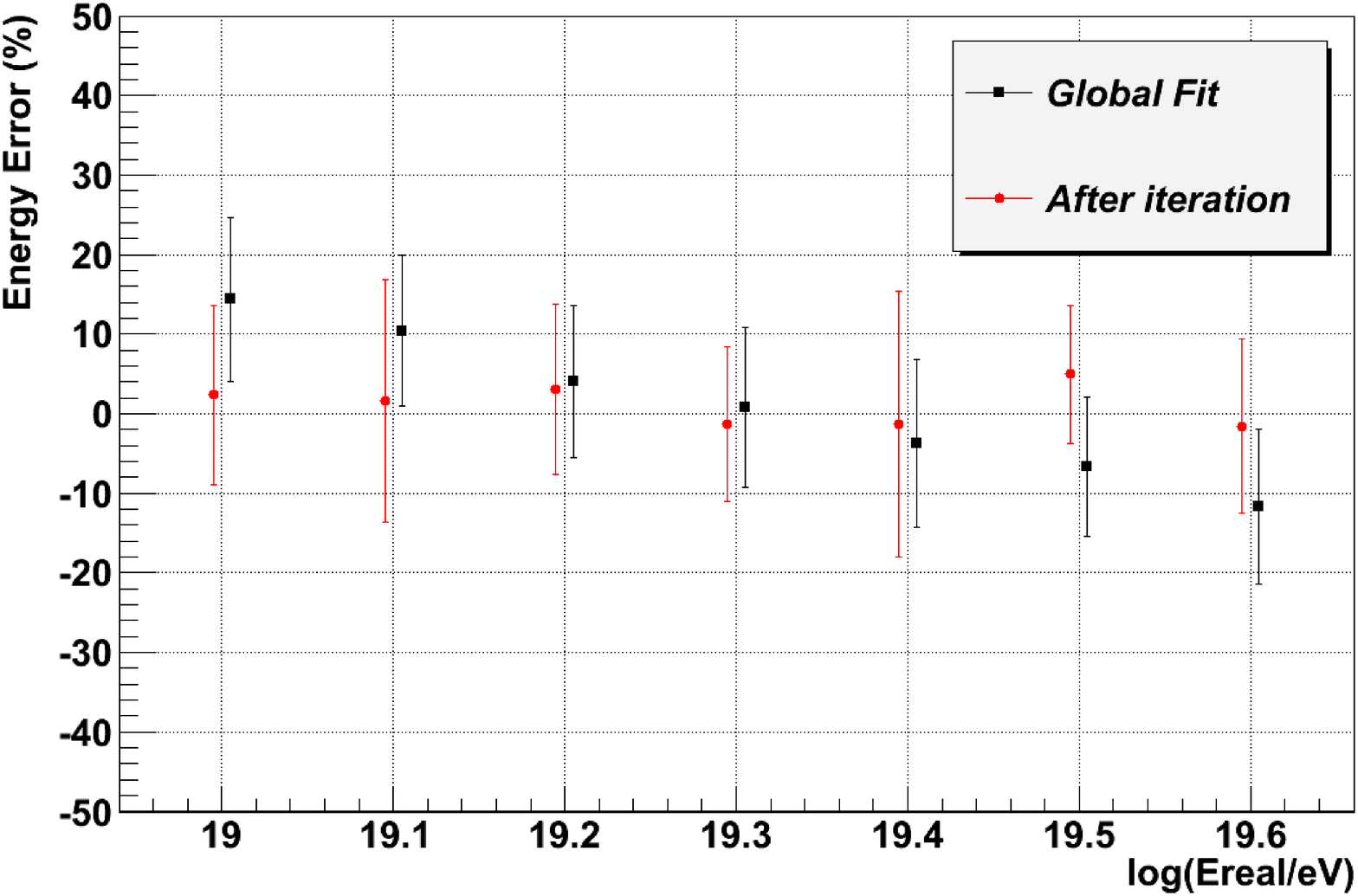}
\end{center}
\caption{Energy error vs. primary energy for proton showers using their own calibration. The points and the error bars are the mean and sigma of the 
Gaussian fit to the energy error distribution respectively. Similar results for iron primaries.}
\end{figure}

In order to account for the muonic component properly, different fits could be performed for each primary energy (Fig. 2
). 
Then, the reconstructed energy obtained with the global fit could be used to select a new calibration curve from Fig. 2
,
repeating the process until the reconstructed energy is not modified. Only one or two iterations are needed. Thus, the energy bias is 
corrected (Fig. 1
, red points) while resolution remains below 12\% considering the whole simulations set.




\begin{figure}[!bt] \label{fig:CalibCurves_Proton}
\begin{center}
\includegraphics[width=8.5cm]{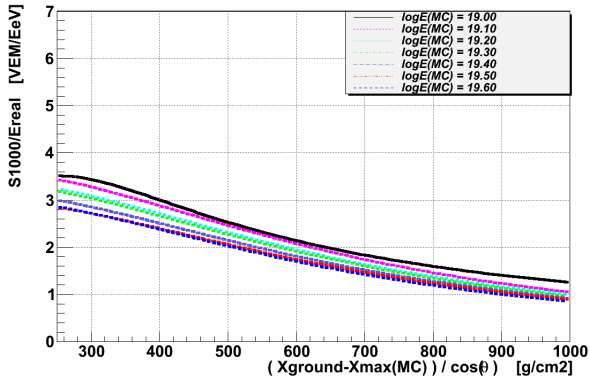}
\end{center}
\caption{$S(1000)/E$ vs $\Delta X$ for proton showers. The fits corresponds to Eq. (\ref{S1000_over_E}). Similar curves are obtained for iron showers.}
\end{figure}

On the other hand, $X_{max}$ is also properly reconstructed. In fact, the bias is negligible and resolution is around 7\% for proton and 4\% for iron 
primaries respectively.




 
Obviously, it is impossible in practice to use different calibration curves for each primary. However, we could also calculate a $mixed$ calibration fitting 
both type of primaries simultaneously (Fig. 3
). The energy ($X_{max}$) error is shown in Fig. 4 
(Fig. 5
). It remains below $10\%$ ($7\%$) for each primary and it is almost negligible if both are considered together. 
Resolution is better than $15\%$ ($7\%$).

\begin{figure}[!bt] \label{fig:CalibCurves_All}
\begin{center}
\includegraphics[width=8.5cm]{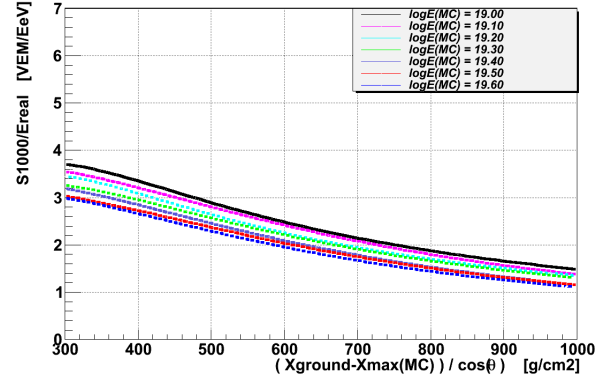}
\end{center}
\caption{As Fig. 2\ref{fig:CalibCurves_Proton} but mixing iron and proton primaries.}
\end{figure}

\begin{figure}[!bt] \label{fig:EnergyBias_vs_logEreal_Mixed}
\begin{center}
\includegraphics[width=8.5cm]{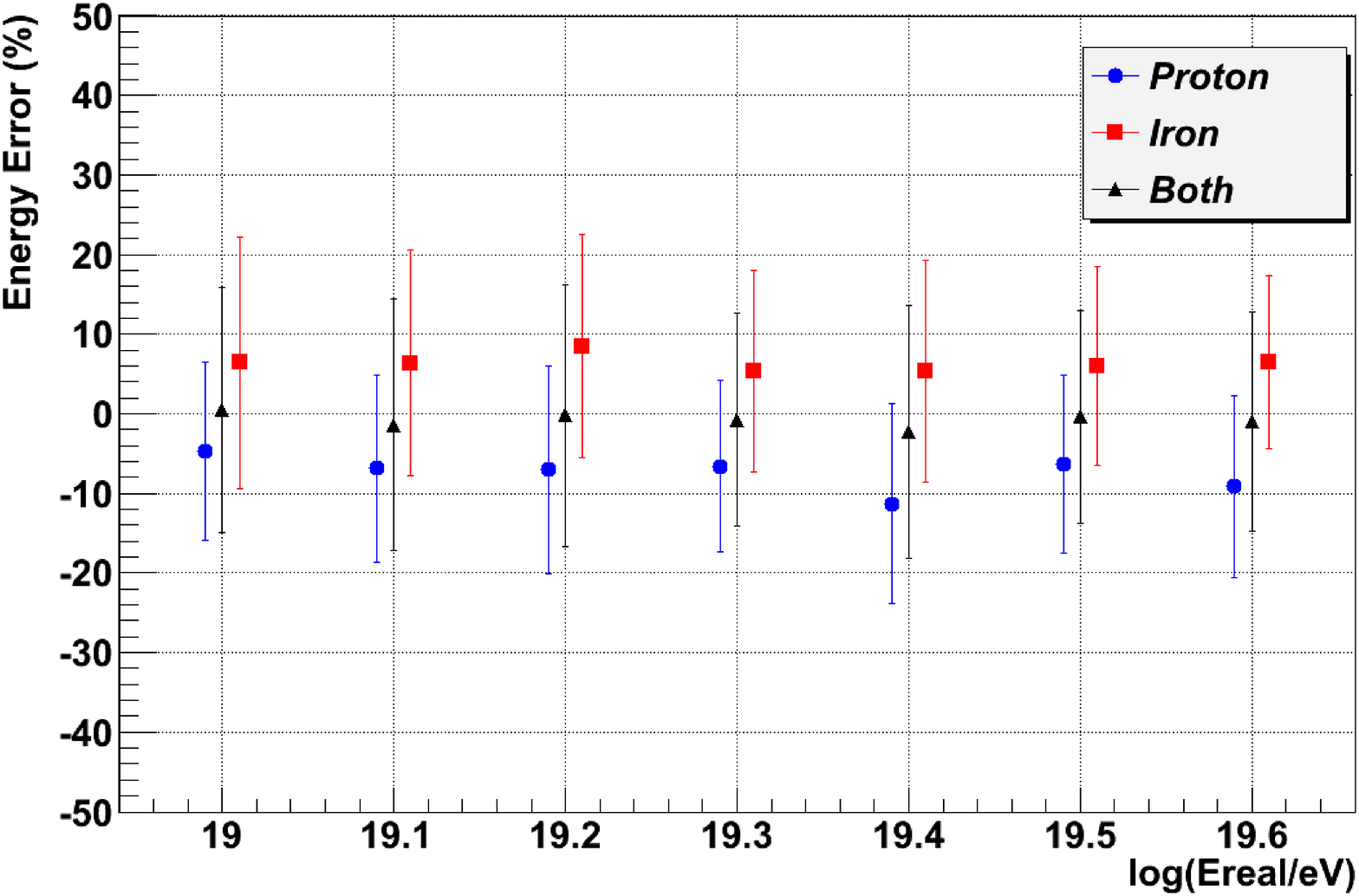}
\end{center}
\caption{Energy error vs. primary energy for proton, iron and both primaries using the $mixed$ calibration. The points and the error bars are the mean and sigma 
of the Gaussian fit to the energy error distribution respectively.}
\end{figure}

\begin{figure}[!bt] \label{fig:XmaxBias_vs_logEreal_Mixed}
\begin{center}
\includegraphics[width=8.5cm]{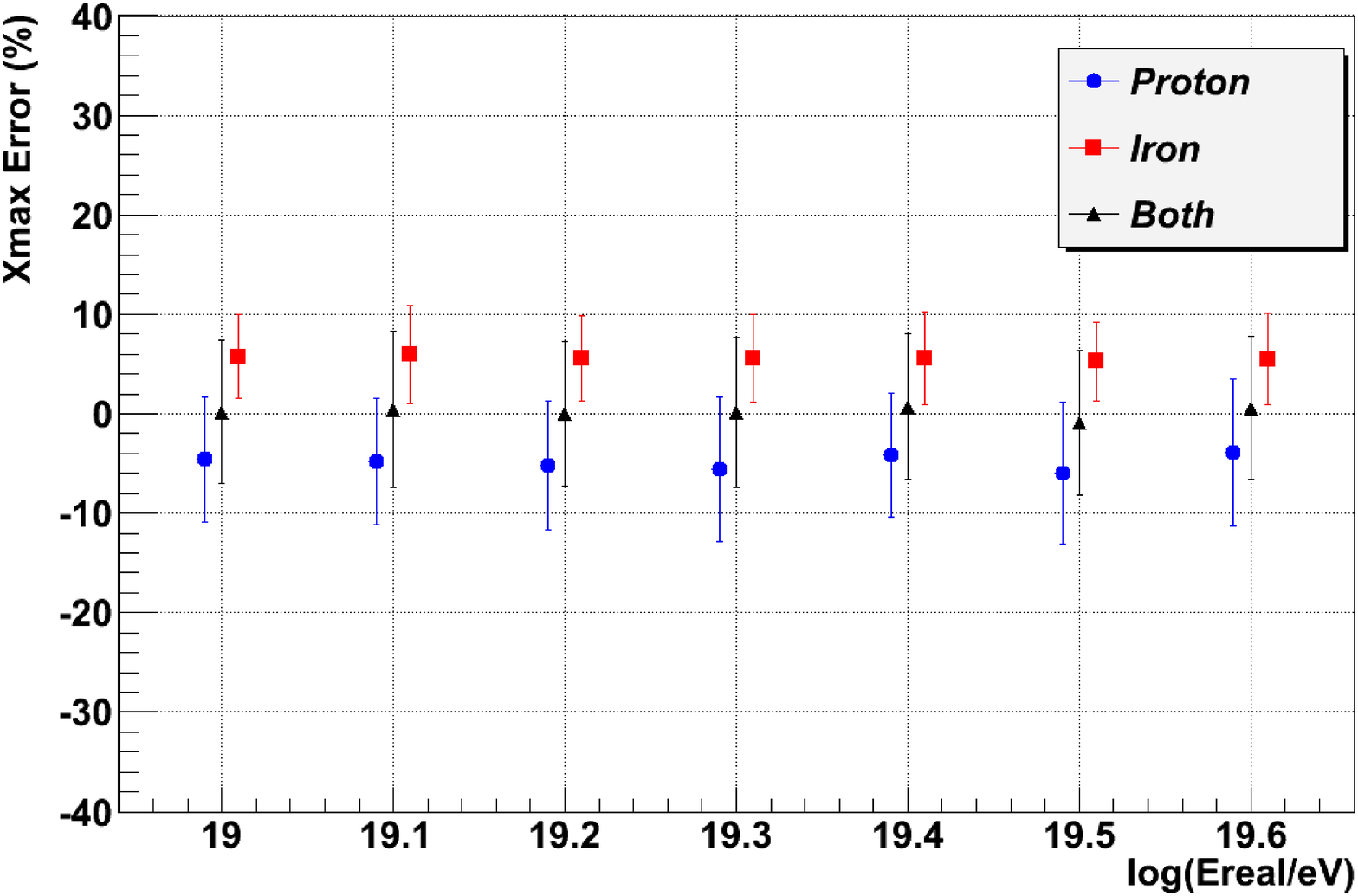}
\end{center}
\caption{As Fig. 4\ref{fig:EnergyBias_vs_logEreal_Mixed} but for $X_{max}$.}
\end{figure}


Finally, we have determined the energy and $X_{max}$ biases in case of a mixed composition sample. We have selected $100$ samples with $100$ events each. Proton and Iron 
primaries are randomly included in the sample such the proton fraction varies from $0$ to $1$ in $0.1$ steps. The energy ($X_{max}$) bias as a function of the proton 
fraction is shown in Fig. 6 
(Fig. 7
). Using the $mixed$ calibration, the energy ($X_{max}$) bias varies 
from $-7\%$ ($-5\%$) to $+6\%$ ($+5\%$) as composition changes from pure proton to pure Iron.

\begin{figure}[!h] \label{fig:EnergyBias_vs_Cp}
\includegraphics[width=8.5cm]{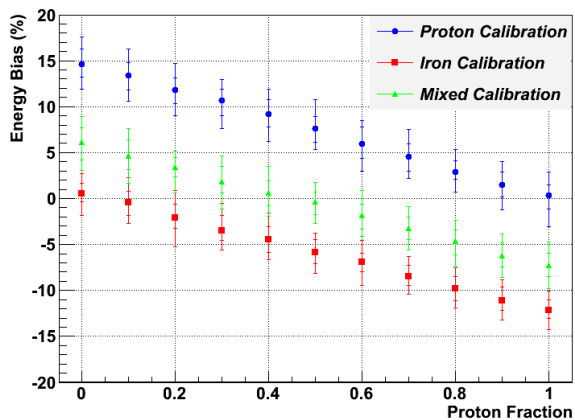}
\caption{Energy bias as a function of the proton fraction using the three sets of calibration curves analyzed in this work. 
The points and the error bars are the median and the confidence levels at $68$ and $95\%$ respectively.}
\end{figure}

\begin{figure}[!h] \label{fig:XmaxBias_vs_Cp}
\includegraphics[width=8.5cm]{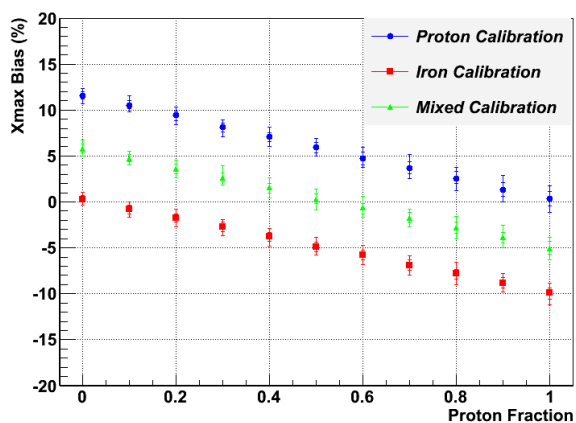}
\caption{As Fig. 6\ref{fig:EnergyBias_vs_Cp} but for $X_{max}$.}
\end{figure}

\section{Conclusions}

An iterative method, previously developed to infer the primary energy of photon-induced showers in pure surface arrays, has been modified to be applicable to hadron-initiated showers. 
The inferred energy bias is negligible and the resolution is around $12\%$ if each primary could be reconstructed with its own calibration. In a more realistic approach, the set of calibration 
curves have been also obtained mixing proton and Iron primaries and then, the bias in the inferred energy varies from $-7\%$ to $+6\%$ as composition changes from pure proton to pure Iron, 
while resolution is better than $15\%$.

In addition, the method allows the indirect determination of the maximum of shower development, $X_{max}$, from pure surface data. The resolution achieved is around $7\%$ and the bias
goes from $-5\%$ for pure proton to $+5\%$ in case of pure Iron. Therefore, $X_{max}$ could be estimated from surface data reliably, whose statistics are around $10$ times larger than that from
fluorescence telescopes. However, it is important to note that MC parametrizations could be affected by the fact that the simulations do not reproduce properly experimental data \cite{Castellina-HIM}.

\section*{Acknowledgments} 

All the authors have greatly benefited from their participation in the Pierre Auger Collaboration. G. R. thanks to Universidad de Alcal\'a (UAH) for a postdoctoral fellowship.
This work is partially supported by Spanish Ministerio de Ciencia e Innovaci\'on (MICINN) under the project FPA2009-11672 and by Mexican PAPIIT-UNAM and CONACyT. Extensive numerical 
simulations were possible by the use of the UNAM super-cluster \emph{Kanbalam} and \textit{SPAS-cluster} at UAH.

\clearpage

\end{document}